\documentclass[a4paper,twocolumn,11pt,accepted=2023-12-27]{quantumarticle}

\pdfoutput=1
\usepackage[final]{graphicx}    
\usepackage{soul}
\usepackage[utf8]{inputenc}
\usepackage[numbers]{natbib}
\usepackage{xcolor}
\usepackage{amsfonts}
\usepackage{amsmath}
\usepackage{amssymb}
\usepackage{hyperref}
\hypersetup{colorlinks=true}

\newcommand{\change}[1]{#1}

\newcommand{\ex}[1]{\mathrm{e}^{#1}}
\newcommand{\Ket}[1]{\left|#1\right>}
\newcommand{\Bra}[1]{\left<#1\right|}
\newcommand{\BraKet}[2]{\left<#1 |#2\right>}
\newcommand{\KetBra}[2]{|#1\rangle\langle#2|}
\newcommand{\trace}[1]{\mathrm{Tr}\left(#1\right)}

\newcommand{\mbf}[1]{\mathbf{#1}}

\newcommand{\erf}[1]{\mathrm{erf}\left(#1\right)}

\begin{document}

\title{Measurement-induced multipartite-entanglement regimes in collective spin systems}

\author{Pablo M. Poggi}
\email{pablo.poggi@strath.ac.uk}
\affiliation{Department of Physics, SUPA and University of Strathclyde, Glasgow G4 0NG, United Kingdom}
\affiliation{Center for Quantum Information and Control, Department of Physics and Astronomy, University of New Mexico, Albuquerque, New Mexico 87131, USA}
\author{Manuel H. Muñoz-Arias}
\affiliation{Institut Quantique and Département de Physique, Université de Sherbrooke, Sherbrooke, Quebec, J1K 2R1, Canada}

\begin{abstract}
We study the competing effects of collective generalized measurements and interaction-induced scrambling in the dynamics of an ensemble of spin-$1/2$ particles at the level of quantum trajectories. This setup can be considered as analogous to the one leading to measurement-induced transitions in quantum circuits. We show that the interplay between collective unitary dynamics and measurements leads to three regimes of the average Quantum Fisher Information (QFI), which is a witness of multipartite entanglement, as a function of the monitoring strength. While both weak and strong measurements lead to extensive QFI density (i.e., individual quantum trajectories yield states displaying Heisenberg scaling), an intermediate regime of classical-like states emerges for all system sizes where the measurement effectively competes with the scrambling dynamics and precludes the development of quantum correlations, leading to sub-Heisenberg-limited states. We characterize these regimes and the \change{crossovers} between them using numerical and analytical tools, and discuss the connections between our findings, entanglement phases in monitored many-body systems, and the quantum-to-classical transition.
\end{abstract}

\date{January 8th, 2024.}
\maketitle

\section{Introduction} 

\begin{figure}[t!]
    \centering
    \includegraphics[width=0.95\linewidth]{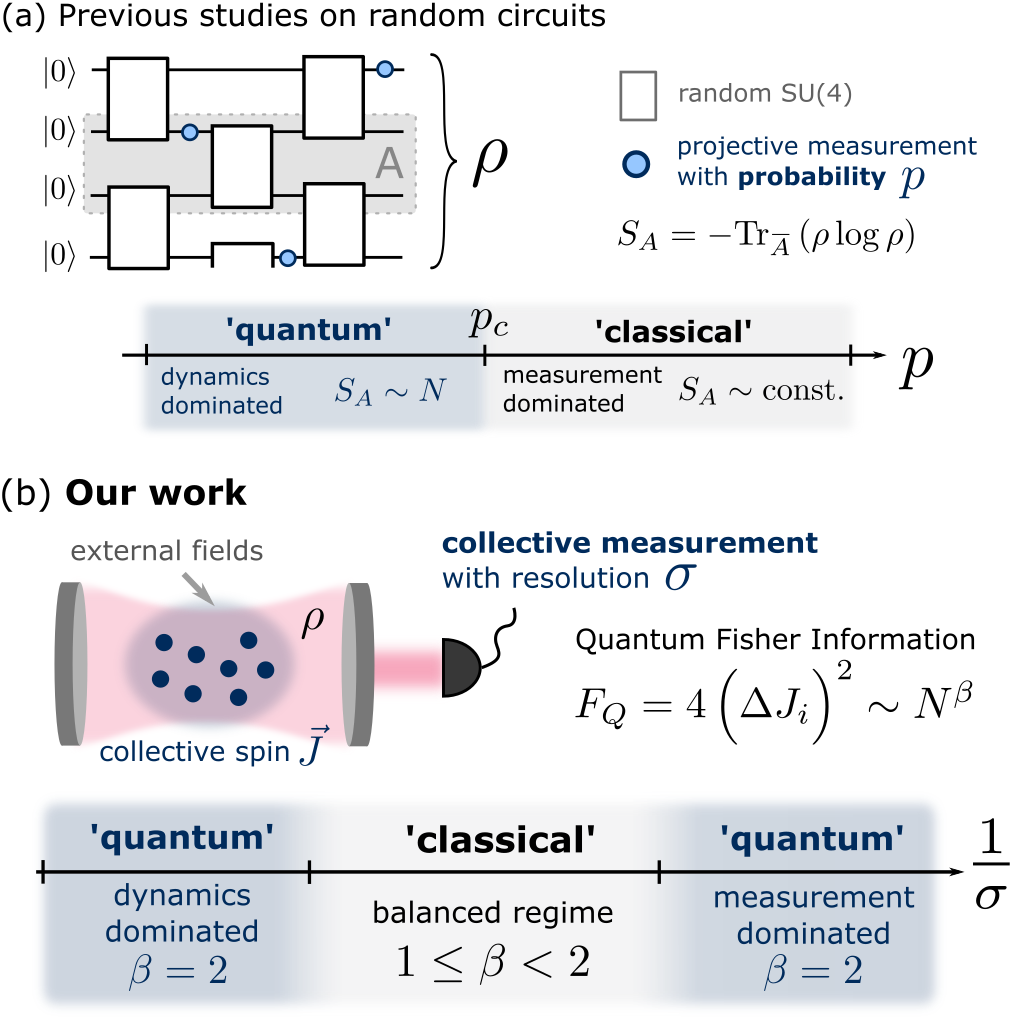}
    \caption{Measurement-transitions in different types of quantum systems. (a) \change{Previous works have studied quantum circuits in presence of measurements which are performed projectively} on single sites and tend to destroy coherence, while unitary dynamics due to interactions tends to generate entanglement. The competition leads to phase transition in the bipartite entanglement as measured by the entanglement entropy $S_A$. The resulting phases can be associated with a dynamics-dominated, highly entangled, `quantum' phase for $p>p_c$, and a measurement-dominated, less entangled, `classical' phase for $p>p_c$.  (b) \change{In our work, we construct a setup analogous to (a) but based on spin systems presenting collective interactions and measurements. In these systems,} both processes tend to generate entanglement. However, an intermediate, classical-like regime appears when the measurement strength $\sigma^{-1}$ is not too strong nor too weak and balances with the scrambling dynamics generated by the unitary interaction. These regimes can be witnessed by the scaling of the Quantum Fisher Information $F_Q$ with system size $N$.}
    \label{fig:figure1}
\end{figure}

The development of quantum technologies like quantum computing and quantum simulation is expected to impact scientific discovery in areas like quantum chemistry, high-energy physics, and condensed matter by enabling efficient simulation of quantum effects in many-body systems~\cite{altman2021,Bauer2023}. A fruitful byproduct of this process is that tools originally developed for quantum information processing are now regarded as new ingredients for models of quantum many-body systems, leading to potentially novel physical phenomena. Examples of this include the study of new interaction models~\cite{piroli2020} and patterns like random graphs~\cite{farhi2022} or multibody interactions~\cite{katz2022} and the characterization of periodic~\cite{else2020} and nonperiodic~\cite{claeys2019} driving in many-body systems. 

A recent development in this context has been the inclusion of quantum measurements intertwined with the unitary dynamics of many-body quantum systems. Efforts towards understanding the competing effects between interactions (which lead to entangled states via quantum information scrambling~\cite{hosur2016}) and projective measurements (which destroy coherence) have led to the discovery of dynamical quantum phase transitions in the properties of these systems~\cite{li2018,skinner2019}. As depicted in Fig.~\ref{fig:figure1} (a), the dynamics of quantum circuits composed of random nearest-neighbour two-qubit gates interleaved with single-site projective measurements happening with probability $p$ displays a sharp transition between a dynamics-dominated phase leading to volume-law entangled states for $p<p_c$ and a measurement-dominated phase leading to area-law entangled states for $p>p_c$~\cite{li2018,skinner2019,bao2020,choi2020,jian2020,gullans2020} \change{(for recent reviews about this phenomenon, see \cite{potter2022,fisher2023})}. Similar transitions have been observed  in the dynamics of systems displaying long-range interactions~\cite{block2022,sierant2022,hashizume2022}, and extensions have been considered exploring the role of various types of measurements, including generalized~\cite{szyniszewski2019} (i.e. nonprojective) and entangling measurements~\cite{van2021}, to the point where similar phenomena can be found without requiring to have unitary dynamics in the process~\cite{ippoliti2021}, or at the level of the non-Hermitian Hamiltonian~\cite{biella2021,gopalakrishnan2021}.

In this work we revisit the role of measurements and their competition with unitary scrambling dynamics in collective spin models, which are systems of spin-$1/2$ particles that interact uniformly among themselves and with external fields and measurement apparatuses~\cite{Stockton2003,lerose2020}. The strict permutational symmetry present in these models allows for their efficient numerical simulation requiring only polynomial resources in the system size $N$. While simple, collective spin models have proven to be excellent toy models for various quantum many-body phenomena, like dynamical~\cite{Corps2022,Corps2023} and excited-state quantum phase transitions~\cite{Cejnar2021}, quantum chaos~\cite{haake1987,munoz2021nonlinear}, dissipative dynamics~\cite{huber2021} and Floquet time crystals~\cite{russomanno2017,munoz2022floquet}, and they are particularly relevant in quantum metrology, where collective all-to-all interactions are used to generate spin-squeezing optimally~\cite{KitagawaUeda1993,Micheli2003,Munoz2023}. Here we will show that these models display different kinds of competing behavior between interactions and collective measurements, which can be diagnosed by its impact on multipartite entanglement and metrological usefulness. 

The key relevant aspect of collective spin systems is that they are measured collectively, which typically generates entanglement among the particles~\cite{Saito2003}. As a consequence, \textit{both} the measurement-dominated dynamics and the interaction-dominated dynamics will lead to highly-entangled states, as schematically depicted in Fig.~\ref{fig:figure1} (b). One would then naively expect that no apparent crossover needs to occur between these regimes. Here we show this is not the case, and that a third regime emerges where the measurement effects are balanced with the unitary dynamics in such a way that the resulting states fail to reach maximum multipartite entanglement. 

We will explore this phenomenon by studying the dynamics of a quantum kicked top (QKT)~\cite{haake1987}, a paradigmatic model of quantum chaos, combined with deterministic generalized measurements of the collective magnetization of the system. These measurements are taken to have variable resolution $\sigma$, which will play an analogous role to the measurement probability $p$ in the random circuits scheme. For these models, we show that for all systems sizes and for generic model parameters three regimes are observed, corresponding to i) measurement-dominated ($\sigma\lesssim 1$), ii) balanced and iii) dynamics dominated ($\sigma\gtrsim N/2)$ behavior, as depicted in Fig.~\ref{fig:figure1} (b) We numerically show that, if the dynamics of unitary model is moderately chaotic, individual quantum trajectories lead to states whose average Quantum Fisher Information (QFI) scales as $N^\beta$, where the maximum $\beta=2$ is obtained for regimes i) and iii), indicating extensive multipartite entanglement and Heisenberg-limited states, while the balanced regime ii) leads to $1\leq \beta <2$, i.e. sub-Heisenberg scaling. For a highly chaotic kicked top, we do not observe signatures of the intermediate regime in the behavior of the scaling exponent, which stays close to $\beta=2$ for all values of $\sigma$. However, we argue that the functional form of the QFI still presents signatures of the existence of a balanced measurement regime.
To show this we propose a toy model based on the action of Kraus operators on Haar-random states which reproduces our numerical results very well, and allows us to show that the observed behavior remains stable for very large system sizes up to $N\sim 10^4$. We further show that the \change{crossover} between regimes is always smooth, and that these monitored collective spin models provide a natural connection between measurement-induced phase transitions observed in many-body models and the fundamental mechanism of the quantum-to-classical transition in systems undergoing weak continuous measurements~\cite{bhattacharya2000}. 

This paper is organized as follows. In Sec.~\ref{sec:collective_systems} we introduce basic aspects of collective spin systems. We discuss the quantum kicked top, which we use as the primary model for unitary dynamics, and describe the types of generalized Gaussian measurements to be implemented. We also define the quantities that will serve as order parameters in our studies and their relation to multipartite entanglement in collective spin systems. In Sec.~\ref{sec:results_transition} we present the results obtained for the hybrid unitary-monitored dynamics at the level of quantum trajectories, and show the existence of the three dynamical regimes as the measurement resolution $\sigma$ is varied. We study the crossovers between those regimes for unitary evolution ranging from completely regular to fully chaotic. In the latter case we present a simplified model which is analytically solvable and accurately reproduces our results. In Sec.~\ref{sec:connections}  we analyze how these results sit on the broader scope of studies on monitored dynamics. We discuss connections with studies on the emergence of classicality in quantum systems undergoing continuous measurements, and analyze the presence of purification transitions in the dynamics of the model. Finally, in Sec.~\ref{sec:outlook} we present our conclusions and lay out potential avenues for future studies.

\section{Scrambling and measurements in collective spin systems} \label{sec:collective_systems}

\subsection{Collective spin models and unitary dynamics}

We consider a system of $N$ spin-$1/2$ particles which interact among themselves and with external agents in such a way that the state of the system is always invariant under permutation of any two particles. Such scenario is most easily described by introducing collective spin operators $J_\alpha=\frac{1}{2}\sum_{i=1}^N \sigma_i^{\alpha}$, $\alpha=x,y,z$. The evolution operator dictating this kind of dynamics commutes with $\mathbf{J}^2=J_x^2+J_y^2+J_z^2$, and the corresponding states belong to the symmetric subspace where $J=N/2$ \change{ and we can write $\mathbf{J}^2=J(J+1)\mathbb{I}$}. A good basis to describe this subspace is given by the Dicke states, i.e. the eigenstates of $J_z$, $\{\Ket{J,m_z}\}$ where $m_z=-J,-J+1,\ldots,J$. The typical initial configuration we will consider is the spin coherent state (SCS) $\Ket{\psi_{\theta,\phi}}=\Ket{\uparrow_{\theta,\phi}}^{\otimes N}$ where all spins point in the same direction. We consider the system to be evolving according to a Floquet map which we write as
\begin{equation}
    U_{\text{KT}}=U_zU_yU_x,
    \label{eq:Uqkt1}
\end{equation}
\noindent where 
\begin{equation}
U_{\mu}=\exp\left[-i\left(\alpha_\mu J_\mu + \frac{k_\mu}{2J} J_\mu^2\right)\right],\ \mu=x,y,z.
\label{eq:Uqkt2}
\end{equation}

This is a variation of the quantum kicked top (QKT) model, an extensively studied model of quantum chaos~\cite{haake1987,kus1987}. The combination of all-to-all interactions (generated by $J_\mu^2\sim \sum_{ij} \sigma_i^\mu \sigma_j^\mu$) with global rotations and time-dependent driving makes this model chaotic for generic choices of $(\alpha_x,\alpha_y,\alpha_z)$ and large values of interaction strengths $(k_x,k_y,k_z)$. In this regime, the iterative application of the map on any SCS generates a highly entangled state which resembles a random state in the symmetric subspace~\cite{trail2008}; we refer to this process as scrambling~\cite{swingle2016,omanakuttan2023}. Because all collective spin systems have a well-defined classical limit in terms of the motion of the collective spin components in a spherical phase space~\cite{Bapst2012}, the transition to chaos in these systems is well understood, and for this particular model it has been studied in previous works~\cite{sieberer2019,omanakuttan2023}.

\subsection{Collective Gaussian measurements} 

Competing with the all-to-all interactions induced by $U_{\text{KT}}$ we will consider generalized collective measurements of a component of the total spin, in this case $J_z$. In the case of an ensemble of atomic spins, this can be regarded as arising from the interaction with a bosonic degree of freedom (like a cavity mode~\cite{deutsch2010quantum}), on which then a Gaussian measurement is performed~\cite{Takahashi1999,Kuzmich2000}. We provide details about this model in Appendix~\ref{app:measurement}. The measurement is described by a continuous set of Kraus operators $\{K_m\}$
\begin{equation}
    K_m = \frac{1}{\left(2\pi \sigma^2\right)^{1/4}}\text{exp}\left(-\frac{(J_z - m)^2}{4\sigma^2}\right),
    \label{eq:kraus_map}
\end{equation}
\noindent  where $\sigma$ is the measurement resolution and $m\in \mathbb{R}$ is the measurement outcome. The associated POVM elements $K_m^\dagger K_m$ obey the completeness relation 
\begin{equation}
    \int\limits_{-\infty}^{\infty} dm\: K^\dagger_m K_m = \mathbb{I},
\end{equation}
\noindent as expected. Given a pure state $\Ket{\psi}$, to model the action of the measurement we sample the probability distribution
\begin{equation}
    P(m) = \Bra{\psi} K^\dagger_m K_m\Ket{\psi},
\end{equation}

\noindent and, for a given outcome $m_0$, construct the post-measured state as
\begin{equation}
    \Ket{\psi_{\mathrm{post}}^{(m_0)}} = \frac{1}{\sqrt{P(m_0)}} K_{m_0} \Ket{\psi},
\end{equation}
\noindent which is already normalized and accounts for measurement backaction. 

In this measurement model, the parameter $\sigma$ can be understood as the width of a Gaussian which sets the resolution of the measurement (alternatively, one can define $\kappa=1/\sigma$ as the measurement strength). As discussed in the previous section and depicted in Fig.~\ref{fig:figure2} (a), different measurement regimes arise depending on the value of $\sigma$. The limit of strong (precise) measurements is obtained when $\sigma\rightarrow 0$. In this case the measurement is close to be projective since 
 \begin{equation}
    P(m)\sim \sum\limits_{m_z=-J}^J |\BraKet{m_z}{\psi}|^2 \delta(m-m_z)
 \end{equation} 
 \noindent and backaction is maximal. The post-measured state approaches $\sim \Ket{J,m_{out}}$, a Dicke state. The opposite limit, when $\sigma\rightarrow \infty$, is that of a  weak (imprecise) measurement, where $P(m)$ is roughly constant and independent of $\Ket{\psi}$. In this regime the measurement outcome carries little information about the state of the system, which is left roughly unchanged due to minimal backaction.
 
\subsection{Entanglement and quantum Fisher information for collective spin states}
\label{ssec:qfi}

Because of the constraints imposed by permutational symmetry, collective spin states possess much less bipartite entanglement (scaling as $\log N$) than the typical many-qubit state~\cite{Stockton2003}. It is thus not expected to observe a transition from area-law to volume-law entanglement in the hybrid dynamics of these systems. Nevertheless, collective spin states lend nicely to be characterized in terms of their multipartite-entanglement properties~\cite{lerose2020}. These can be quantified via the Quantum Fisher Information (QFI)~\cite{Pezze2018}:
\begin{equation}
    F_Q[J_{\alpha},\Ket{\psi}]\equiv 4 \left(\Delta J_{\alpha}\right)^2=4\left(\Bra{\psi}J_{\alpha}^2\Ket{\psi} - \Bra{\psi}J_{\alpha}\Ket{\psi}^2\right).
    \label{eq:qfi_def}
\end{equation}

For \change{product} states the QFI satisfies $F_Q\leq N$ and the equality is saturated by spin coherent states \change{$\Ket{\uparrow_{\mathbf{n}}}^{\otimes N}$ such that $\mathbf{n}$ is perpendicular to the direction of $\alpha$}.  If $F_Q>N$, then the state is entangled, and more specifically the condition
\begin{equation}
    F_Q[J_{\alpha},\Ket{\psi}] > Nk
    \label{eq:multi_entang_condition}
\end{equation}
\noindent implies that there are at least $k$ particles entangled in the system~\cite{pezze2009}. The ultimate bound of the QFI is thus $F_Q\leq N^2$. Often the study of this quantity is done in terms of the QFI density $f_Q\equiv F_Q/N$ and distinguishes regimes where $f_Q$ is intensive ($f_Q\sim \mathrm{const.}$) or extensive ($f_Q\sim N$) in its dependence on system size $N$. 


Crucially, the QFI also determines the ultimate sensitivity achievable in a quantum sensing protocol where a parameter $\theta$ is encoded into the system as $e^{i\theta J_{\alpha}}\Ket{\psi}$. More specifically, the variance of an unbiased estimator of $\theta$ is lower bounded by $F_Q^{-1}$~\cite{Braunstein1994}. Thus, the QFI of Eq.~(\ref{eq:qfi_def}) connects the level of multipartite entanglement present in a pure state $\Ket{\psi}$ to its usefulness as a quantum sensing probe. Product states achieve at best a sensitivity of $N^{-1}$ (standard quantum limit), while use of entangled states can improve the scaling of the variance to be $N^{-2}$ (Heisenberg limit)~\cite{Pezze2018}.

Instead of focusing on the QFI with a specific choice of direction as in Eq.~(\ref{eq:qfi_def}), here we will consider the QFI averaged over three orthogonal directions
\begin{equation}
    \overline{F_Q}[\Ket{\psi}]=\frac{1}{3}\sum\limits_{\alpha=x,y,z} F_Q[\Ket{\psi},J_{\alpha}]
    \label{eq:avg_qfi}
\end{equation}
\noindent which is independent of the particular choice of axis {(see Appendix~\ref{app:mean_qfi}). We refer to this quantity as the mean QFI, for short. For permutationally symmetric states we have that $J_x^2+J_y^2+J_z^2=J(J+1)\mathbb{I}$ and so Eq.~(\ref{eq:avg_qfi}) reduces to
\begin{equation}
    \overline{F_Q}[\Ket{\psi}]=\frac{4}{3}\left[J(J+1)-||\langle \mathbf{J}\rangle||^2\right],
\end{equation}
\noindent from which we see that when the length squared of the magnetization vector $||\langle \mathbf{J}\rangle||^2=\sum_{\alpha}\langle J_{\alpha}\rangle^2$ can be used to witness the level of multipartite entanglement in the state $\Ket{\psi}$.  This is because a mean QFI exceeding $Nk$ implies that the QFI corresponding to at least one component of $\mathbf{J}$ will satisfy the condition in Eq.~(\ref{eq:multi_entang_condition}). \change{See \cite{hyllus2012} for a detailed analysis on $\overline{F_Q}$ as a multipartite entanglement witness}.

The mean QFI can be readily computed for several states of interest which will serve as a reference for our analysis. In Table \ref{tabl:avgqfis} we show the values of $\overline{F_Q}$ obtained for the case of Haar random (symmetric) states, randomly-picked Dicke states, spin squeezed states and spin coherent states,  which are derived in Appendix \ref{app:mean_qfi}. Recalling that $J=N/2$, one can read off the scaling properties for each case. As noted before spin coherent states are not entangled and thus lead to $\overline{F_Q}\sim N$. Spin-squeezed states are entangled and lead to enhanced scaling with $N$ which is not universal. On the other hand, both Haar and Dicke states display, on average, $\overline{F_Q}\sim N^2$ and thus show extensive multipartite entanglement. For large system sizes, Haar states reach slightly higher mean QFI ($\sim N^2/3$) than typical Dicke states ($\sim 2 N^2/9$).

\begin{table}

\begin{tabular}{ c|c } 
 State type & Mean QFI $\overline{F_Q}$ \\ 
 \hline
 Spin coherent states (SCS) & $\frac{4}{3} J$  \\
 Spin squeezed states &  $\frac{4}{3}J\cosh(r(J))$  \\
 Haar-random states (average)& $\frac{4}{3}\left(J^2 + \frac{J}{2}\right)$  \\ 
 Dicke states (average) & $\frac{8}{9}\left(J^2+J\right)$  \\
\end{tabular}

\caption{Values of the mean QFI $\overline{F_Q}$ defined in Eq.~(\ref{eq:avg_qfi}) for specific state types. The scaling of this quantity with system size $J=N/2$ is linear for unentangled states like spin coherent states, and quadratic for highly entangled states like random Haar or Dicke states.}
\label{tabl:avgqfis}
\end{table}

\section{Competition regimes between measurements and scrambling dynamics} \label{sec:results_transition}

In order to characterize the hybrid unitary-monitored dynamics in these systems, we study the behavior of individual quantum trajectories constructed in the following way. Starting from a spin coherent state along a randomly chosen direction $\Ket{\psi_0}=\Ket{\psi_{\theta,\phi}}$, we iteratively apply a unitary map $U_{\mathrm{KT}}$ followed by a measurement described by Eq.~(\ref{eq:kraus_map}). Thus, at each step the map takes the form
\begin{equation}
    \Ket{\psi_{j+1}}=\frac{1}{\sqrt{P(m_j)}} K_{m_j} U_{\mathrm{KT}}\Ket{\psi_j},
    \label{eq:dyn_model_map}
\end{equation}

\noindent where
\begin{equation}
    P(m) = \Bra{\psi_j}U_{\mathrm{KT}}^\dagger K_m^2 U_{\mathrm{KT}}\Ket{\psi_j}.
    \label{eq:dyn_model_distro}
\end{equation}

The unitary part is dictated by the kicked top map $U_{\mathrm{KT}}$ of Eq.~(\ref{eq:Uqkt1}). We choose the rotation parameters to be $\alpha_x = 1.7$, $\alpha_y=1$ and $\alpha_z=0.8$, and the twisting coefficients as $k_x = 0.85k$, $k_y = 0.9k$ and $k_z = k$. This choice is inspired by previous works~\cite{sieberer2019,omanakuttan2023} and guarantees that the QKT model is fully regular for $k=0$ (corresponding to just rotations of the collective magnetization vector), has a regular regime for small $k\lesssim 1$, and becomes fully chaotic for $k\gtrsim 2.5$. 

For a given initial state, each trajectory is labeled by a set of measurement outcomes $\mathbf{m}={m_0,m_1,\ldots,m_{M_s}}$ where $M_s$  is the number of iteration steps. Following Sec.~\ref{ssec:qfi}, for each trajectory a time-average of the length of the mean magnetization vector $||\langle \mathbf{J}\rangle||^2$ is computed, and the mean QFI defined in Eq.~(\ref{eq:avg_qfi}) is stored as the relevant order parameter \footnote{\change{We point out that the classical Fisher Information (FI) has been used in \cite{bao2020} to characterize entanglement transitions in monitored random circuits. In that study, the FI measures how much classical information about the initial state of the system can be extracted from the measurement outcomes. As such, the definition of the FI depends on the type of measurement. In our work, we treat the QFI as a property of the state which witnesses multipartite entanglement. This definition of the QFI is thus independent of the kind of dynamics we study.}}. Since the evolution is stochastic, the procedure is repeated $M_{\mathrm{runs}}\sim 50$ times, giving good convergence. Results reported in what follows are averages over these random instances. In most cases the standard error of the mean with respect to this average is small and so for clarity we do not plot error bars. However, we present detailed plots including such error bars in Appendix~\ref{app:additional_numerics}, \change{where we also present additional details on the numerical methods}.\\

\begin{figure}[t!]
    \centering
    \includegraphics[width=1\linewidth]{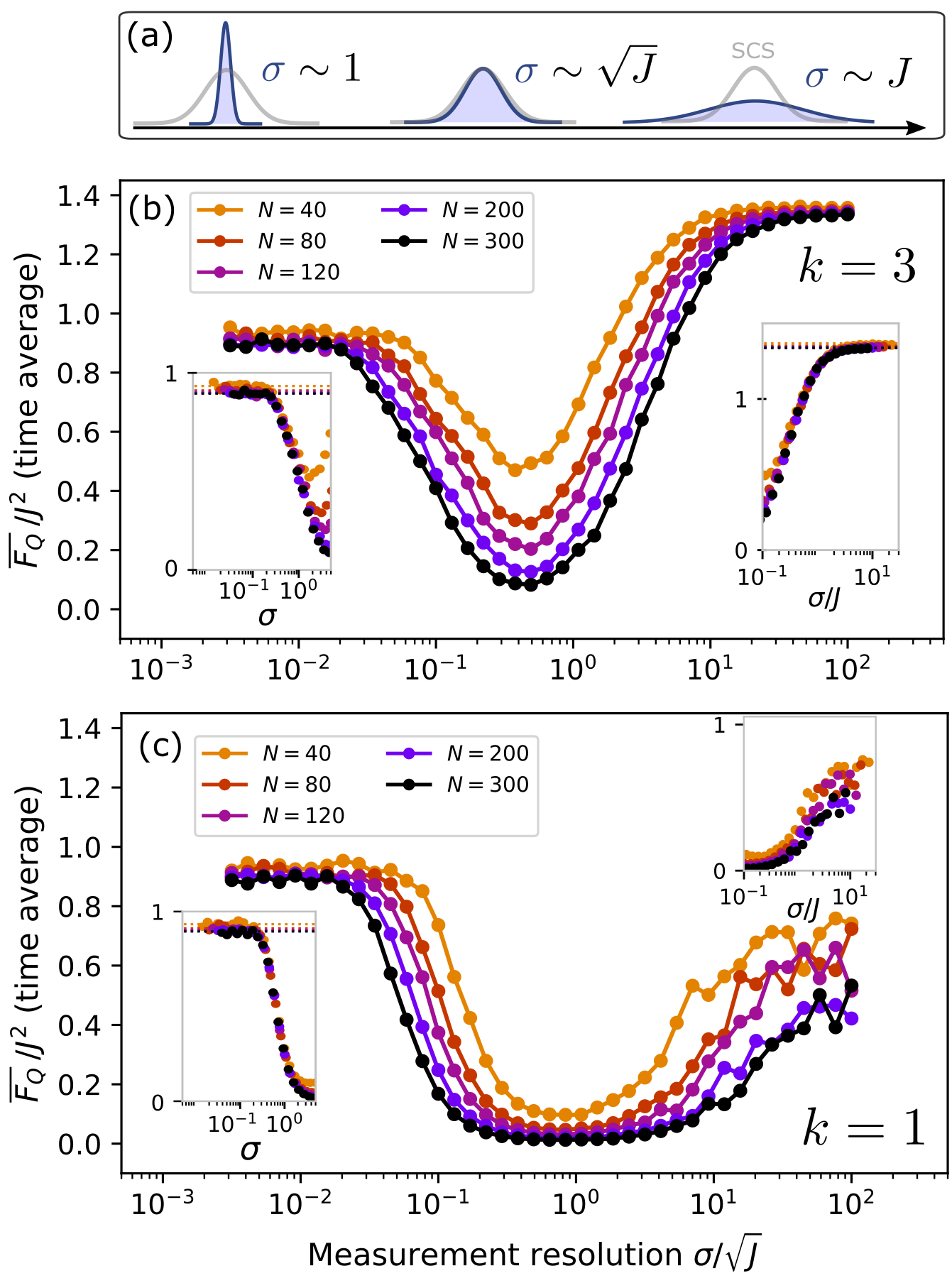}
    \caption{(a) Schematic depiction of the three measurement regimes referenced to in the main text. The Gaussian-like curves represent both the measurement of Eq.~(\ref{eq:kraus_map}), parameterized by $\sigma$ (in blue) and the typical distribution of a spin coherent state (SCS) in the Dicke basis (in gray). (b) and (c) Mean QFI $\overline{F_Q}$ computed from quantum trajectories evolved following the hybrid dynamical map in Eq. (\ref{eq:dyn_model_map}). Each point corresponds to a fixed choice of system size $N$ and measurement resolution $\sigma$, and is obtained after averaging over time, choice of initial state and measurement record. The number of iterations of the map is $M_s=40$ in all cases. Results are shown as a function of $\sigma/\sqrt{J}$ in the main plot, as a function of $\sigma$ in the leftmost insets, and as a function of $\sigma/J$ in the rightmost insets. In the insets, the dashed lines correspond to the mean QFI values expected for typical Dicke states (left) and Haar-random states (right), which can be found in Table \ref{tabl:avgqfis}. Plot (b) and (c) correspond to a choice of $k=3$ (chaotic) and $k=1$ (regular) in the KT map, respectively. See main text for more details. }
    \label{fig:figure2}
\end{figure}

In Fig.~\ref{fig:figure2} (b) and (c) we show numerical results of the time-averaged mean QFI $\overline{F_Q}/J^2$ for various system sizes $N=2J$ as a function of the measurement resolution $\sigma/\sqrt{J}$ . As schematically depicted in Fig.~\ref{fig:figure2} (a), this choice of normalization intends to measure $\sigma$ in units of the typical projection noise of a spin coherent state. Panels (b) and (c) correspond to $k=3$ and $k=1$, such that the QKT model is chaotic and regular, respectively. The behavior of the order parameter displays three distinguishable regimes depending on the magnitude of the measurement resolution $\sigma$. Small $\sigma/\sqrt{J}$ is the \textit{strong} measurement regime where backaction dominates the dynamics. At each time step, the state is roughly projected back to a Dicke state $\Ket{J,m_z}$, but the value of $m_z$ is essentially random as it depends on the application of the unitary $U_{\mathrm{KT}}$ which continuously changes $\langle J_z\rangle$. The left inset in each plot shows the same data as the main plot but only for small values of the measurement resolution and plotted as function of $\sigma$ unscaled. We see that the data collapses into a single curve which is independent of $N$ and starts at roughly $\sigma\sim 1$. This is the minimum relevant resolution of the problem (independently of $N$) and it corresponds to the measurement being able to resolve individual collective spin eigenstates (i.e. individual values of $m_z$). As a consequence, lowering the measurement resolution further yields no change in the dynamics and the curves lead to a constant value which is given by the mean QFI of an average Dicke state shown in Table \ref{tabl:avgqfis}, $\overline{F_Q}\simeq 8J(J+1)/9$ which scales as $N^2$ and is indicated with dotted lines in the inset. Thus, the states generated by the hybrid unitary-monitored dynamics display extensive multipartite entanglement and Heisenberg scaling on average. Notice that this behavior is seen in both the case of chaotic KT dynamics, shown in Fig.~\ref{fig:figure2}(b) and regular KT dynamics, shown in Fig.~\ref{fig:figure2}(c), which indicates that this regime is \textit{measurement dominated} and thus independent of the details of the unitary part of hybrid evolution. \\

The opposite regime is that of \textit{weak} measurements at large $\sigma/\sqrt{J}$. In these cases measurement backaction is minimal at each step, and the hybrid evolution is dominated by the unitary contribution. The right inset of Fig.~\ref{fig:figure2}(b) and (c) shows the data plotted as a function of $\sigma/J$. For the chaotic case (a) we see that all curves coincide around $\sigma\sim J$, which is now the \textit{largest} relevant resolution of the problem, corresponding to a case where the measurement is unable to extract any meaningful information about $J_z$, whose spread in eigenvalues is $\sim 2J$. As a result, further increasing $\sigma$ has no effect on the dynamics, and the mean QFI tends to the Haar-random value (see Table \ref{tabl:avgqfis}) $\overline{F_Q} \simeq 4J(J+1/2)/3\sim N^2$ and thus displays Heisenberg scaling and extensive multipartite entanglement. For $k=1$ (when the QKT is mostly regular) the behavior of the mean QFI is not generic and the states fail to achieve the Haar-random limit.

In between these two cases, an intermediate regime emerges when $\sigma \sim \sqrt{J}$, where the mean QFI presents a noticeable dip in its value. The minimum happens roughly at the same value of $\sigma/\sqrt{J}$ for all curves, indicating that this choice of normalization for the measurement resolution is the correct one, and complementary to the limiting cases discussed above, c.f. Fig.~\ref{fig:figure2} (a). Importantly, we observe that $\overline{F_Q}/J^2$ becomes dependent on the value of $J$ in this regime, which signals a change in scaling with system size. This is particularly noticeable for $k=3$ but also present in most other cases as long as $k$ is not too large (such cases will be studied in more detail in Section~\ref{ssec:fully_chaotic}). We point out that a full set of data is shown in Appendix~\ref{app:additional_numerics}, where we include other values of $k$. 

To further investigate of each of these regimes, we numerically study the scaling properties of the resulting states as a function of system size. To achieve this, for each fixed value of $\sigma/\sqrt{J}$ we fit the data to an ansatz $\overline{F_Q}\ = c N^\beta$. We repeat this procedure for several choices of the QKT parameter $k$. The resulting exponents $\beta(k)$ are plotted in Fig. \ref{fig:figure3} (a) as a function of $\sigma/\sqrt{J}$. There, we observe that Heisenberg-limited states (i.e. $\beta\simeq 2$) are achieved for strong measurements independently of $k$, which is expected as in this regime measurement backaction dominates over the unitary contribution of the hybrid evolution. Apart from this regime, \change{whether Heisenberg scaling is achieved or not} depends on the value of $k$. For weak measurements, this is again expected, since  Haar-random states arise naturally in the long-time dynamics of the QKT (which is only weakly perturbed by the measurement) as long as we work in the chaotic regime $k\gtrsim 2$. In the intermediate regime, however, an interesting behavior arises: even for cases where QKT dynamics is chaotic, the resulting states display sub-Heisenberg scaling, with exponents that in some cases are closer to the standard quantum limit $\beta=1$. This is seen in the moderately chaotic regime when $2\lesssim k \lesssim 5$. In this regime, the measurement competes with the scrambling induced by the unitary dynamics by reducing the projection noise, but only to the level of a spin coherent state. Conversely, the unitary dynamics rotates the state at each step, effectively leading to a subsequent measurement of a different spin component at each instance. This, in turn, precludes the measurement from having a cumulative effect over time, which would lead effectively to a strong measurement and a projection onto a Dicke state. We thus conclude that in this intermediate regime a balance is achieved between measurement and unitary dynamics, with the resulting dynamical behavior being explained only by the combination of their effects. Finally, we observe that when the QKT is in the strongly chaotic regime $k\gtrsim 7$, the curves tend to flatten and be close to $\beta\sim 2$ for all measurement strengths. In this regime, which will be studied in more detail in Sec. \ref{ssec:fully_chaotic}, each application of the QKT creates an approximately random state in such a way that measurements of moderate strength are unable to compete to reduced the overall uncertainty of the state. We point out that this finding is reminiscent to a recent result in Ref.~\cite{morral2023}, where it was observed that highly chaotic quantum circuits competing with unentangling unitaries (instead of measurements) can lead to an stable phase of volume-law-entangled states.

\begin{figure}[t!]
    \centering
    \includegraphics[width=1\linewidth]{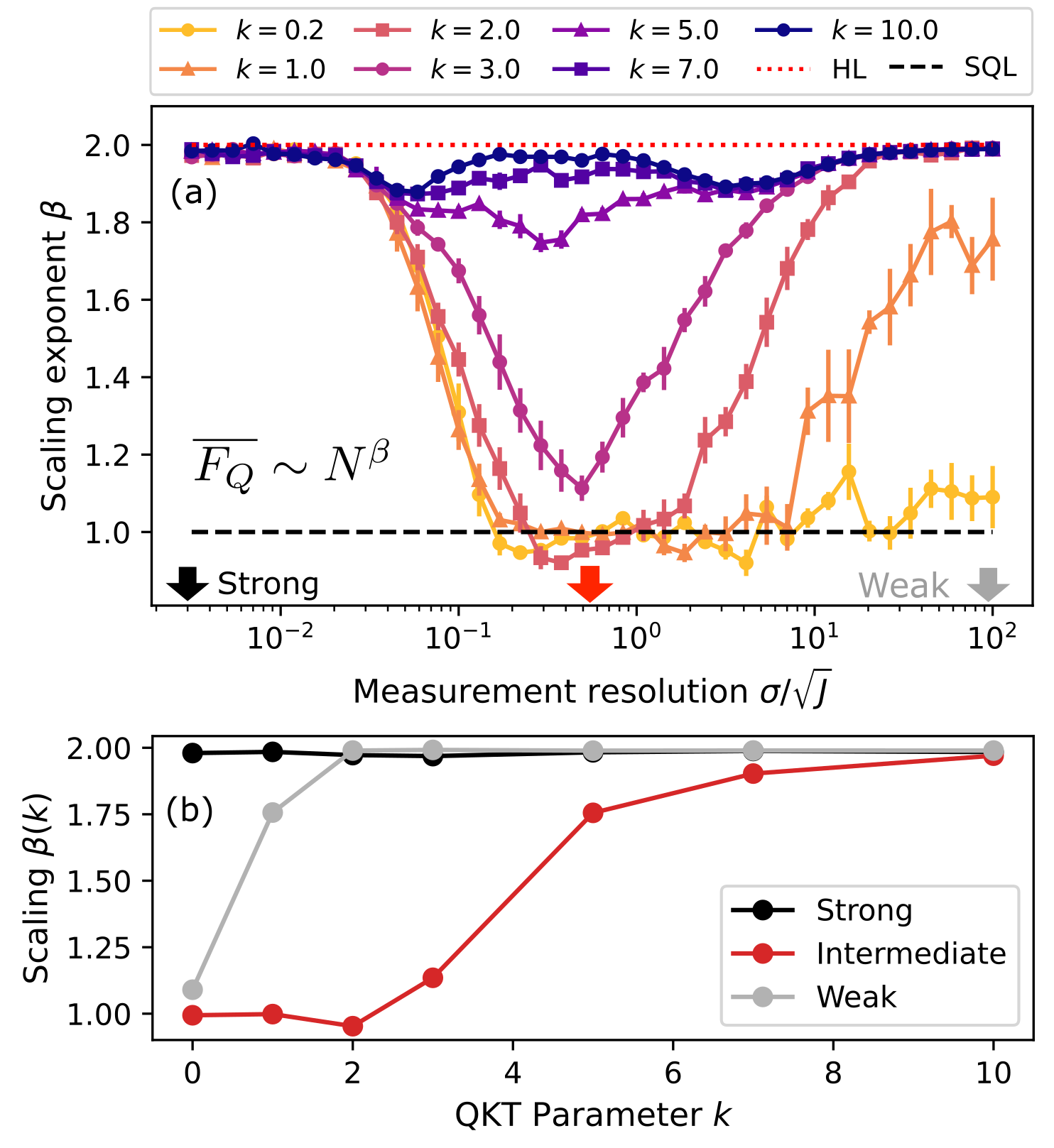}
    \caption{ Scaling of the mean QFI with system size $\overline{F_Q}\sim N^{\beta}$. Figure (a) shows the scaling exponent $\beta$ as a function of the rescaled measurement resolution $\sigma/J$ for different values of the KT parameter $k$, which changes the unitary dynamics from fully regular ($k\lesssim 1$), to mixed, to fully chaotic for ($k\gtrsim 2.5$). Plot (b) displays the same exponents as a function of $k$ for three representative values of the measurement resolution, corresponding to the strong, intermediate and weak measurement regimes, which are indicated as arrows in the horizontal axis of (a). }
    \label{fig:figure3}
\end{figure}

This overall behavior can be further rationalized by analyzing the scaling exponent as a function of the QKT parameter $k$. In Fig.~\ref{fig:figure3} (b) we plot $\beta(k)$ as a function of $k$ for three representative values of $\sigma/\sqrt{J}$, indicated with arrows in Fig.~\ref{fig:figure3} (a) and corresponding to the three relevant measurement regimes. 
The case of weak measurements (gray curve) can be readily understood by the regular-to-chaotic transition in the QKT dynamics, showing that as $k$ increases Haar-random states (displaying Heisenberg scaling) are achieved. Similarly, the strong-measurement regime curve (in black) shows that the scaling is independent of $k$ and fixed at $\beta\simeq 2$, as expected for Dicke states on average. However, the intermediate regime case (red curve) shows that fully scrambling dynamics leading to Heisenberg-limited states is recovered only when $k\gtrsim 7$. We thus observe a shift of the transition to chaos towards higher values of $k$. This is caused by the collective measurements which, in this regime, inhibits the generation of long-range correlations in phase space (and thus the generation of random states). We will revisit this phenomenon when we discuss the connection to the quantum-to-classical transition in Sec.~\ref{sec:connections}.

\subsection{Fully chaotic case} \label{ssec:fully_chaotic}

The results of Fig.~\ref{fig:figure3} (b) indicate that when the QKT parameter $k$ is large enough, the states generated by the hybrid dynamics always display $\overline{F_Q}\sim N^2$ independently of the measurement strength and the crossover between different scalings is lost. However, we show here that the intermediate regime can still be identified in the functional form of the mean QFI. In Fig.~\ref{fig:figure4} we show numerical results for $N=300$ and $k=10$ (gray triangles) where such regime can be observed as a dip in the value of the mean QFI. In order to explore this phenomenon further, we studied a simplified model for the dynamics inspired by the fact that, for highly chaotic evolution, a single application of the QKT map yields a random-like state irrespective of the initial condition. We thus expect to be able to describe this case by studying the action of the Kraus map of Eq.~(\ref{eq:kraus_map}) on Haar-random states $\Ket{\psi}$,
\begin{equation}
    \Ket{\Psi_{m}} = \frac{1}{\sqrt{P(m)}} K_{m} \Ket{\psi}
\end{equation}

\noindent the resulting configuration being a function of the measurement resolution $\sigma$ and the random measurement outcome labeled $m$. As noted in Eq. (\ref{eq:avg_qfi}), the mean QFI of the resulting state can be computed from the length of the magnetization vector $||\langle \mathbf{J}\rangle||^2$ and thus only depends on quantities of the form 
\begin{equation}    \Bra{\Psi_{m}}J_{\alpha}\Ket{\Psi_{m}}^2 = \frac{\Bra{\psi} K_{m}^\dagger J_{\alpha} K_{m}\Ket{\psi}^2}{\Bra{\psi} K_{m}^\dagger  K_{m}\Ket{\psi}^2}.
    \label{eq:haar_expec}
\end{equation}

\begin{figure}[t!]
    \centering
    \includegraphics[width=1\linewidth]{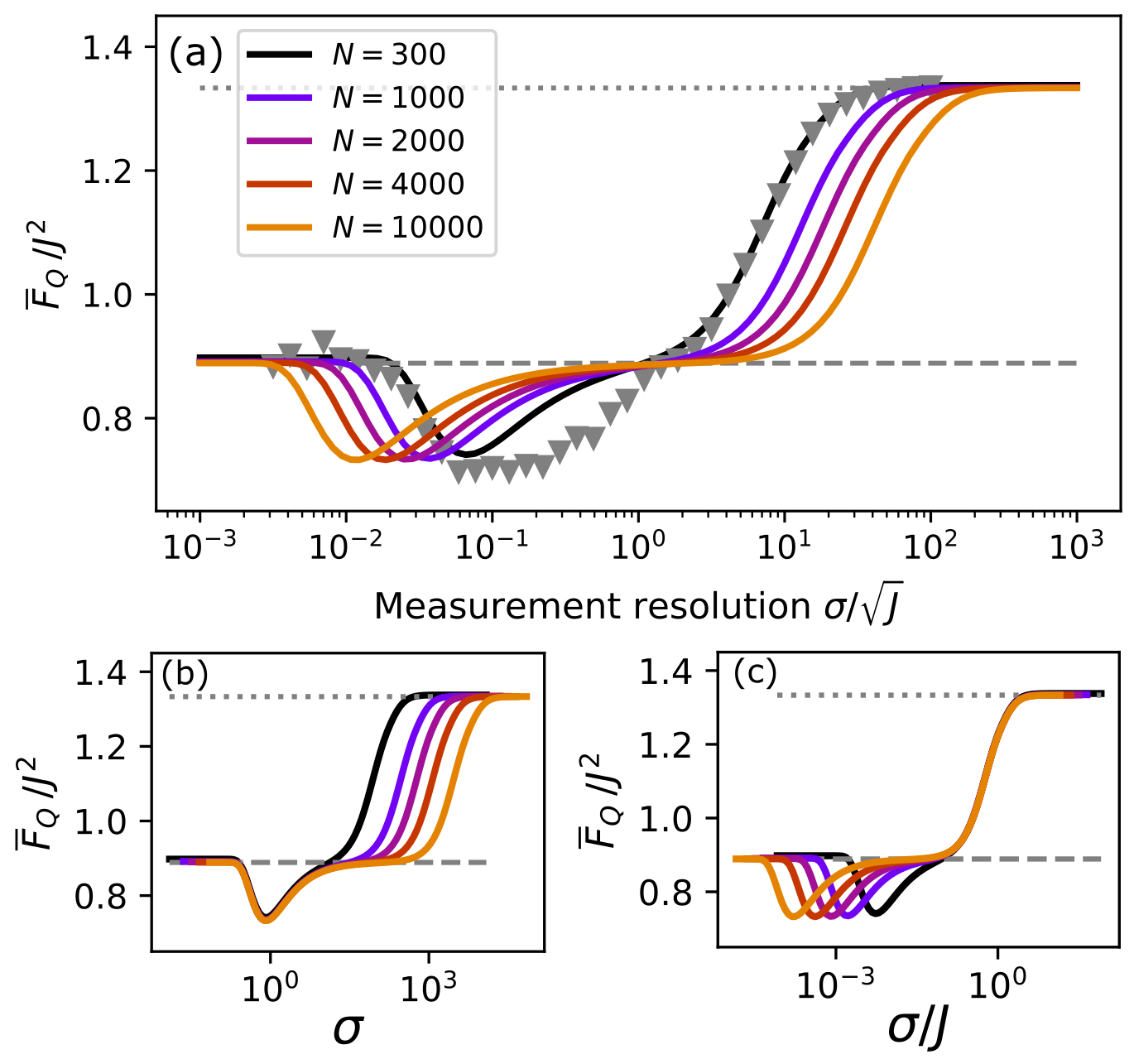}
    \caption{Mean QFI resulting from the analytically-solvable model described in Sec. \ref{ssec:fully_chaotic}. Full curves are obtained from the expressions in Eq. (\ref{eq:haar_resultJ}), whose closed form is given in Appendix \ref{app:analytical_model}, plotted for various system sizes $N$. For the case of $N=300$, we show numerical results (gray triangles) obtained in a similar way as those shown in Fig. \ref{fig:figure2}. Similar curves are shown in all three panels, plotted as a function of (a) $\sigma/\sqrt{J}$, (b) $\sigma$ and (c) $\sigma/J$. The dotted and dashed lines correspond to $\overline{F_Q}/J^2=4/3$ and $\overline{F_Q}/J^2=8/9$, which are the high$-J$ limits of the random Haar and Dicke states of Table \ref{tabl:avgqfis}. }
    \label{fig:figure4}
\end{figure}

We then aim at integrating Eq.~(\ref{eq:haar_expec}) over the uniform (Haar) distribution of states for all three collective spin components $J_{\alpha}$, $\alpha=x,y,z$, and use Eq.~(\ref{eq:avg_qfi}) to construct the mean QFI $\overline{F_Q}$ for this simplified model. We find that the integral of the ratio in Eq.~(\ref{eq:haar_expec}) is accurately estimated by the ratio of the integrals, as expected by the approximation $\mathbb{E}[x/y]\simeq \mathbb{E}[x]/\mathbb{E}[y]$ valid as long as $\mathbb{E}[y]>0$ and not too small~\cite{duris2018} (here $\mathbb{E}[\cdot]$ symbolizes the average over the Haar distribution). So, we can use the known result
\begin{equation}
    \int\limits_{\rm Haar} d\psi \Bra{\psi}O\Ket{\psi}^2 = \frac{1}{d(d+1)}\left( \trace{O}^2 + \trace{O^2}\right)
\end{equation}
\noindent valid for any operator $O$ (see~\cite{collins2006} and Appendix A of~\cite{poggi2020}), to estimate 
\begin{equation}
    \mathbb{E}\left[\Bra{\Psi_{m}}J_{\alpha}\Ket{\Psi_{m}}^2\right] \simeq \frac{ \trace{K_{m}^2 J_{\alpha}}^2 + \trace{K_{m}^2 J_{\alpha} K_{m}^2 J_{\alpha}} }{\trace{K_{m}^2}^2 + \trace{K_{m}^4}}.
    \label{eq:haar_resultJ}
\end{equation}

In Appendix~\ref{app:analytical_model} we show that closed analytical expressions can be obtained for these quantities in the limit $N\gg 1$. The resulting expressions have a complicated form and are not very transparent, and so we omit presenting them in the main text. However, having access to these analytical expressions where $N$ is a parameter means we can calculate numerically the mean QFI for large system sizes. In order to compare with data as that of Fig.~\ref{fig:figure2}, we compute numerical averages of Eq.~(\ref{eq:haar_resultJ}) over the measurement outcome $m$, and plot the results as a function of $\sigma$ for different values of $N$ in Fig.~\ref{fig:figure4} (a). In the figure we show cases ranging from $N=300$ up to $N=10^4$. For the former, we can compare with the numerical simulations, where we observe excellent agreement in the strong and weak measurement regimes. While the intermediate regime shows less quantitative agreement, we find that the overall shape and magnitude of the mean QFI are indeed captured by the simplified model (notice the difference of vertical scales between Fig.~\ref{fig:figure2} (a) and~\ref{fig:figure4} (a)). 

For the remaining cases, we observe the curves show similar shapes as the system size is increased and the weak and strong measurement cases achieve the values expected from Table~\ref{tabl:avgqfis}. In a similar way as we observed in the data of Fig.~\ref{fig:figure2}, plotting the curves in different scales reveals the limiting regimes in a clean way. Fig.~\ref{fig:figure4} (b) shows the mean QFI as a function of the unscaled measurement resolution $\sigma$, where we observe that for $\sigma\lesssim 1$ all curves coincide indicating the onset of the strong measurement regime. An analogous behavior is observed in Fig.~\ref{fig:figure4} (c) as a function of $\sigma/J$, where the curves coincide for $\sigma \gtrsim J$ as the model describes a weak measurement situation. In between these regimes, we find that the unifying feature is a node at $\sigma/\sqrt{J}\simeq 1$ where all curves seem to intersect. Interestingly these are not present in the `moderately' chaotic regime studied at the beginning of this section, but we argue that it can be regarded as a qualitative indicator of an intermediate regime where the measurement tends to counteract the action of the unitary dynamics. The results of this model suggest that such regime will exist for all system sizes, and thus that the two crossovers depicted in Fig. \ref{fig:figure1} (b) remain in place in the thermodynamic limit. 

\section{Discussion and connections to previous results} \label{sec:connections}

\subsection{Quantum-to-classical transition in continuously monitored quantum systems}

Measurements play a key role in the study of fundamental aspects of quantum theory, and particularly in  elucidating the underlying mechanisms for the quantum-to-classical transition, i.e. the process whereby quantum features such as interference and superposition fade away and lead to an effective classical description. This phenomenon has been famously described by the mechanism of environmentally-induced decoherence~\cite{paz2002,schlosshauer2007}, where the system interacts with an external agent (the environment) which is inaccessible to the observer. This process, however, fails to describe the emergence of classical \textit{trajectories}, which are ultimately the fundamental object describing classical mechanics. 

This problem was noted by Bhattacharya, Habib and Jacobs in Ref.~\cite{bhattacharya2000}, who studied the emergence of classicality in the quantum trajectories arising when a quantum system undergoes weak continuous measurements.  The authors show that in a regime where the measurement strength is large enough to produce localization of a state in phase space (e.g. by reducing the uncertainty in position of a single particle) but also weak enough such that the dynamics is not overpowered by noisy measurement backaction, then the ensuing dynamics of the quantum trajectories recovers the appropriate classical behavior.

The regime of not-too-strong but not-too-weak measurements described in Ref.~\cite{bhattacharya2000} is precisely the one discussed in the previous sections of this work, which is schematically depicted in Fig.~\ref{fig:figure2} (a). Indeed, our findings illustrate that measurements of moderate strength can make the state of the system localized in all spin directions. On the other hand, strong measurements unbalance the variances due to the uncertainty principle (as happens in the regime of small $\sigma$ leading to Dicke states, as seen in Fig.~\ref{fig:figure2}), and weak measurements fail to suppress the development of long-range correlations in phase space created by the unitary chaotic dynamics (of the QKT model in our case). 

We emphasize that collective spin systems can be fully described in phase space (via their Wigner or Husimi distributions, see for instance~\cite{Takahashi1976,Polkovnikov2010}) where typically the association $\hbar\sim N^{-1}$ is made, and thus the notion of measurement-induced localization alluded to by Bhattacharya \textit{et al.} has a precise meaning for these systems as well. In our calculations, however, we have used quantum information-theoretic measures like the QFI to analyze this behavior. Our results show that the crossover between regimes are smooth, with the classical behavior recovered continuously as the measurement strength approaches the optimal value $\sigma\sim \sqrt{J} \sim 1/\sqrt{\hbar}$. This smooth recovery of the classical limit was also observed in collective spin systems undergoing continuous measurements and feedback~\cite{munoz2020,munoz2020_pspin}.

These observations lead us to state that the intermediate measurement regime described in Sec.~\ref{sec:results_transition} is associated to an emergent classical behavior of the system, which aligns with the fundamental mechanism described in Ref.~\cite{bhattacharya2000} but which differs from classical regimes found in the context of measurement-induced phase transitions with local monitoring. In such cases, the classical phase is typically attributed to a regime where the measurement dominates and the most amount of classical information is retrieved from the system. This is not the case for our model since measurement-dominated dynamics still leads to entangled (nonclassical) states and delocalization in phase space. Thus, we conclude that the mechanism explaining the emergence of a classical regime in monitored many-body systems depends on the properties of the measurement model under consideration, and in particular that collective measurements can lead to a novel kind of mechanism for the emergence of a classical regime, as compared to the case of local monitoring.

\subsection{\change{Multipartite entanglement transitions in other monitored spin systems}}

\change{The QFI has been studied recently in Ref.~\cite{paviglianiti2023} in the dynamics of the integrable transverse-field Ising model combined with local measurements. In such a system, the dynamics of the bipartite entanglement entropy shows a measurement-induced phase transition. When considering the dynamics at the level of the non-Hermitian Hamiltonian (i.e. in the no-click limit), the authors find that the QFI density shows a transition between intensive and extensive at a critical measurement rate that matches the one observed for the usual bipartite entanglement transition. However, when considering the full stochastic dynamics (averaged over quantum trajectories), the QFI interestingly shows a distinctive behavior where a third regime appears at low measurement rates, in which the QFI density becomes intensive again. While the nature of the measurements is markedly different, it is intriguing to note that the QFI acts as a good indicator of three dynamical phases for the system in Ref. \cite{paviglianiti2023} as well as in our case. Notice that while both spin models are also quite different, they both present global symmetries which distinguishes them from generic scrambling many-body systems: the QKT is a collective spin system with strict permutational symmetry, while the Ising model is integrable by mapping to free fermions (see also \cite{loio2023}) for the role of this symmetry in measurmement-induced transitions).}

\subsection{Purification transitions}

In the analysis presented in Sec.~\ref{sec:results_transition} we used as an order parameter the mean QFI of the individual quantum trajectories generated by the map in Eq.~(\ref{eq:dyn_model_map}). This is because this quantity can witness the degree of entanglement of the states, and so it represents a natural analog to the bipartite entanglement entropy considered in the context of quantum circuits. However, recent results have established that measurement-induced phase transitions can also be observed in the behavior of the purity of the states at the level of quantum trajectories~\cite{gullans2020}, provided one initializes the system in a mixed state $\rho_0$. Previous studies have observed a similar transition as the one depicted in Fig.~\ref{fig:figure1} (a) but seen in the average purity of the state generated by the hybrid dynamics, such that for small measurement probability $p<p_c$, the measurement fails to reduced the entropy of the state significantly and the state remains highly mixed. Conversely, for frequent measurements $p>p_c$ the state is purified due to the measurement backaction. The analogy between measurement-induced entanglement and purification transitions has been widely observed for models of local measurements, and they are often regarded to be two manifestations of the same phenomenon~\cite{block2022,hashizume2022,sierant2022,gopalakrishnan2021}. It is thus natural to explore whether this is also the case in the model of collective measurements studied here.

To tackle this question, we numerically simulate the dynamics of the map in Eq.~(\ref{eq:dyn_model_map}) starting from a maximally mixed symmetric state $\rho_0=\mathbb{I}/d$, $d=N+1$. For mixed states this map takes the form
\begin{equation}
    \rho_{j+1}=\frac{1}{P(m_j)} K_{m_j} U_{\mathrm{KT}}\rho_j U_{\mathrm{KT}}^\dagger K_{m_j}^\dagger.
    \label{eq:dyn_model_map2}
\end{equation}

We study the purity of the state $\eta(\rho)=\trace{\rho^2}$ time-averaged, and ensemble-averaged over quantum trajectories, as a function of system size $N=2J$ and measurement resolution $\sigma$. Results are shown in Fig.~\ref{fig:figure5} for (a) $k=3$, (b) $k=1$ and (c) $k=10$, which are the same values of $k$ we have used when analyzing the behavior of the mean QFI and the multipartite entanglement in the previous section. 

The results indicate that the two limiting cases of small and large $\sigma$ recover the expected behavior. First, weak measurements disturb the state only mildly, and since the unitary part of the map cannot change the purity, the initial mixed state stays mixed throughout the evolution. On the other hand, strong measurements tend purify the state and after a few iterations of the map the state is completely pure. These cases have analog regimes in the entanglement properties observed in Fig.~\ref{fig:figure2}. Crucially, however, we observe no apparent intermediate regime in the purification dynamics, as the purity seems to merely changes from low to high as the measurement gets stronger. This is further confirmed by the fact that the data collapses to a single curve when plotted as a function of $\sigma$ for \textit{all} measurement regimes, as seen in the inset of Fig.~\ref{fig:figure5} (a). Recall that, for the entanglement analysis, the region of collapse changed depending on how the horizontal axis was scaled. When visualized as a function of $\sigma/\sqrt{J}$, no salient behavior is observed in the results, indicating that the balanced regime observed for the entanglement when $\sigma \sim \sqrt{J}$ is not present in the purification dynamics. 

Finally, we also observe that changing the level of chaoticity of the unitary kicked top map by varying $k$ renders only minor changes in purity. This is in stark contrast with the entanglement behavior, particularly at large $\sigma$, where the unitary part of the map dominates the dynamics. For this case, however, this is not surprising since a maximally mixed state is invariant under unitary dynamics (irrespective of its level of chaoticity). Note that for the entanglement analysis, chaotic dynamics was essential to delocalize the initial state, which in turn generated a competition with the measurement for moderate measurement strength. This is clearly no longer the case for initial mixed states, which are highly delocalized irrespective of the unitary dynamics. 
We thus conclude that, for this model, the purification and entanglement regimes and their corresponding crossovers are markedly different. This is thus another distinctive feature introduced by the collective measurements with respect to the local measurement case. 

\begin{figure}[t!]
    \centering
    \includegraphics[width=1\linewidth]{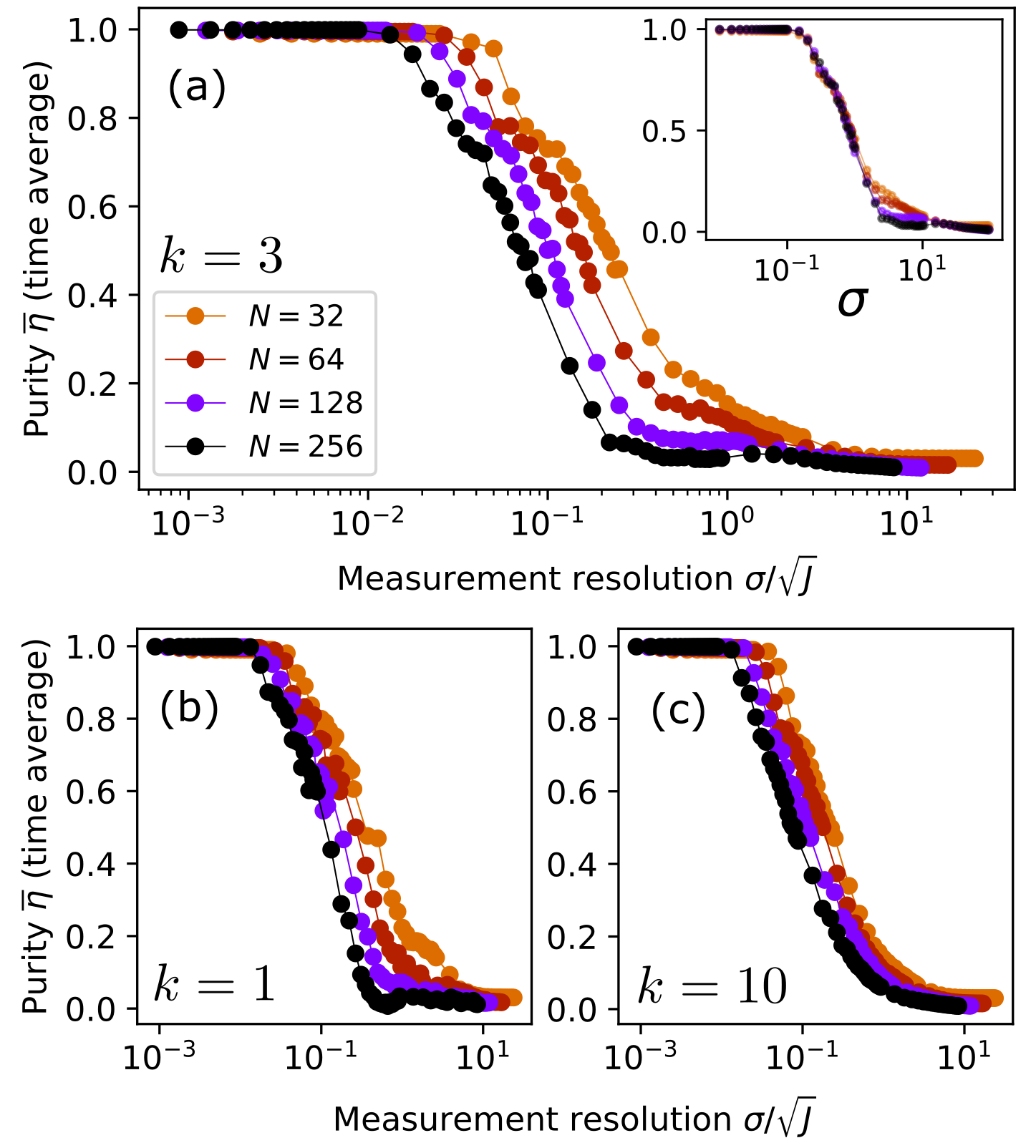}
    \caption{Purification \change{crossover} in the dynamics of the hybrid map of Eq.~(\ref{eq:dyn_model_map2}). All plots show the behavior of the time- and ensemble-averaged purity of the individual quantum trajectories after $3N$ applications of the map, as a function of the measurement resolution. Each plot corresponds to a different value of the kicked top nonlinearity parameter $k$: (a) $k=3$, (b) $k=1$ and (c) $k=10$. The inset of (a) shows the same data as the main plot but plotted as a function of the unscaled $\sigma$ to illustrate the data collapse. Similar behavior is seen for all values of $k$, although the explicit plots are not shown here. }
    \label{fig:figure5}
\end{figure}




\section{Outlook} \label{sec:outlook}
We have studied the dynamics of spin systems undergoing collective unitary dynamics interleaved with collective measurements. While seemingly analogous to the setup of measurement-induced phase transitions studied in quantum circuits, we found that the proposed model displays three dynamical regimes as a function of the measurement strength, instead of two. These regimes lead to highly entangled states in the limiting cases of strong and weak measurements, but also to a balanced, intermediate regime where the resulting states are less entangled and, thus, more classical. Because of the symmetry properties of collective spin systems, we find that the quantum Fisher information (QFI), related to multipartite entanglement, is a good order parameter to describe these regimes, and in particular the scaling exponent of the mean QFI with system size $\overline{F_Q}\sim N^{\beta}$ is a useful numerical metric. We find that the scaling varies from $\beta \sim 2$ in the limit cases (indicating Heisenberg-limited states) to $1 < \beta < 2$ in the intermediate regime (indicating sub-Heisenberg limited states). The \change{crossover} between regimes is smooth as a function of measurement resolution $\sigma$, thus precluding the existence of phase transitions. We found that the crossover in scaling is present as long as the unitary part of the dynamics is not in the highly chaotic regime; however we showed that clear signatures of the three regimes can still be observed and even analytically studied if that condition is not met. 

The results we have obtained show that the model proposed here introduces distinctive features which are not present in the usual case of local measurements. We have further illustrated this by showing that our results are deeply connected  to fundamental aspects of the quantum-to-classical transition as described by the seminal work of Bhattacharya, Habib and Jacobs~\cite{bhattacharya2000}. \change{This connection allows to argue that the classical regime arising in our study is a manifestation of a general mechanism which arises naturally when studying monitored quantum dynamics in phase space}.
Finally, we have studied the behavior of the purity in the dynamics of the model and showed that these regimes are not in one-to-one correlation to the ones seen in the behavior of the multipartite entanglement.


We note that the model introduced in this work combines different elements which have been studied separately in previous works related to measurement-induced phase transitions. For instance, the effect of long-range interactions in the unitary part of the dynamics was studied in \cite{block2022,sierant2022}. In many of these cases it was observed that the volume-law phase is suppressed by the presence of long-range interactions approaching the all-to-all limit \change{where the system is permutationally symmetric} (there are exceptions, for instance when considering fast scramblers with non-local couplings, which actually make the volume-law phase more stable \cite{hashizume2022}). In our case, the all-to-all symmetric interaction pattern of the QKT means that the state of the system is always in a permutationally symmetric configuration which does not support states with extensive bipartite entanglement. Therefore, an interesting avenue for future work is to explore the effects of unitary dynamics with no symmetric constraints (e.g. by using all-to-all random couplings, or sparse graphs) in combination with collective measurements. If states with extensive bipartite entanglement can be reached, then actual phase transitions might occur between the regimes described in Sec.~\ref{sec:results_transition}.

Similarly, the effect of entangling measurements has also been studied in \cite{szyniszewski2019,van2021,ippoliti2021}, and the overall finding is that entangling measurements can also be the driving force behind the emergence of entangled and unentangled regimes in monitored dynamics. In our case, however, we have found that a novel intermediate regime can be found where the entangling effects of the dynamics and the measurements balance each other out. It is an outstanding question to reveal how collective the measurements have to be to recover this behavior in more general models. 

Finally, measurement-induced transitions are typically regarded as challenging to be observed experimentally~\cite{noel2022,hoke2023}. This is because of the stochastic nature of the monitored evolution, leading to a large overhead in the amount of experiments that need to be performed~\cite{moghaddam2023}, and also of the fact that order parameters like the entanglement entropy are complex to measure in the lab. We point out that the order parameter we have considered here is quite simple, since the mean QFI is constructed from one- and two-particle correlation functions \change{(however, it does still require the large overhead in post-selecting measurement outcomes)}. Furthermore in certain physical platforms like ultracold atoms in optical cavities~\cite{muniz2020,li2022}, collective monitoring is a natural model to consider. However, in order to assess how feasible it is to observe the regimes studied here experimentally, a detailed analysis of the effects of dissipation and decoherence, which are intrinsic to the atom-light interface setting, should be performed~\cite{deutsch2010quantum,Baragiola2014}.

\acknowledgements
We are grateful to Jun Takahashi for insightful discussions about the nature of phase transitions in monitored quantum systems, and to Philip Blocher and Shane Dooley for useful comments on the first version of this manuscript. Work at the University of Strathclyde was supported by AFOSR grant number FA9550-18-1-0064. This material is partially based upon work supported by the U.S. Department of Energy, Office of Science, National Quantum Information Science Research Centers, Quantum Systems Accelerator (QSA). Additional support is acknowledge from the Canada First Research Excellence Fund.

\appendix

\section{Measurement model} \label{app:measurement}

Here we briefly describe the origin of the measurement model of Eq.~(\ref{eq:kraus_map}) as arising from the interaction, in the dispersive regime, between a propagating linearly polarized probe beam and the pseudo spin of a collection of two-level atoms which has been initially magnetized. In particular, we consider the context of a Faraday rotation experiment~\cite{deutsch2010quantum,Baragiola2014}. 

In this type of experiment the direction of propagation of the light defines the $z$-direction. The polarization degree of freedom is described in a quantized form via its position on the Poincar\'e sphere. Using the Schwinger representation for a two-mode oscillator, we write the components of the Stokes vector as 
\begin{align}
    S_1 &= \frac{1}{2} (a^{\dagger}_H a_H - a^{\dagger}_V a_V), \\
    S_2 &= \frac{1}{2} (a^{\dagger}_H a_V + a^{\dagger}_V a_H), \\
    S_3 &= \frac{1}{2i} (a^{\dagger}_H a_V - a^{\dagger}_V a_H), 
\end{align}
where $a_{H,V}^\dagger$ ($a_{H,V}$) are creation (annihilation) bosonic operators corresponding to Horizontal and Vertical polarizations of the light, respectively. In terms of the collective atomic spin $\vec{J}$ and the Stokes vector of the probe light $\vec{S}$, the interacting Hamiltonian can be written as~\cite{deutsch2010quantum}

\begin{equation}
    H = \frac{\chi}{\Delta t} J_z\otimes S_3\ \rightarrow\ U(\Delta t)=\exp{-i\chi J_z\otimes S_3},
\end{equation}

\noindent where $\chi$ is the Faraday rotation angle per unit angular momentum. Notice that, if the input probe beam is linearly polarized, say along the $x$-direction, the Stokes vector is initially pointing in the positive $x$-direction of the Poincar\'e sphere, and the above interaction rotates it by a small angle such that the final position lies on the equator and somewhere in between the positive $x$ and positive $y$ directions. As such, the interaction has transfer a small amount of population into the $V$ mode. This small change can be detected by using a polarimeter set to the $S_2$ direction (diagonal / antidiagonal).\\

To describe the measurement, we assume the initial state of the probe to be a two-mode coherent state which is polarized along the horizontal direction. This state can be written as $\Ket{\alpha_H,0_V}$. For this state, $\langle \vec{S}\rangle=(N_L/2,0,0)$. If $|\alpha_H|^2\equiv N_L \gg 1$, we can resort to the Holstein-Primakoff approximation~\cite{Holstein1940} for the probe light and define
\begin{eqnarray}
X_L&=&\frac{1}{\sqrt{N_L/2}}S_2 \\
P_L&=&\frac{1}{\sqrt{N_L/2}}S_3
\end{eqnarray}

\noindent such that $[X_L,P_L]\simeq i$. For the state mentioned above is easy to see that $\langle X_L\rangle = \langle P_L\rangle = 0$ and $\langle X_L^2\rangle = \langle P_L^2\rangle = \frac{1}{2}$ and thus we approximate it by the vacuum state $\Ket{0_L}$ in the Holstein-Primakoff plane. Notice that in these new variables Faraday rotation corresponds to an $x$-displacement in the Holstein-Primakoff plane proportional to the atomic magnetization. 

If the state of the ensemble is initially $\Ket{\psi_A}$, then after the interaction we have

\begin{equation}
    \rho_{AL}(\Delta t) = U(\Delta t) \KetBra{\psi_A,0_L}{\psi_A,0_L} U^\dagger(\Delta t)
\end{equation}

After measuring the probe in the $S_2=\sqrt{N_L/2}X_L$ basis we obtain the (random) outcome $x_L$, and then the post-measured atomic state is (apart from normalization)

\begin{equation}
    \Bra{x_L}\rho(\Delta t)\Ket{x_L}=K_{x_L} \KetBra{\psi_A}{\psi_A} K_{x_L}^\dagger
\end{equation}

\noindent where we have introduced the Kraus operator $K_{x_L}$ describing the measurement. This operator has the form 
\begin{align}
    K_{x_L} &= \Bra{x_L} \ex{-i \chi \sqrt{\frac{N_L}{2}}J_z \otimes P_L}\Ket{0_L} \\
    &= \sum\limits_{m_z} \BraKet{x_L - \sqrt{\frac{N_L}{2}} m_z}{0_L} \KetBra{m_z}{m_z} \\
    &= \exp\left[-\frac{1}{2}\left(x_L - \chi \sqrt{\frac{N_L}{2}} J_z\right)^2\right]
    \end{align}

After rearranging the terms in the exponent, defining $m = \sqrt{\frac{2}{\chi^2 N_L}} x_L$ and $\sigma^2 = \frac{1}{N_L \chi^2}$, and adding the appropriate normalization factor, we obtain

\begin{equation}
    K_m = \frac{1}{\left(2\pi \sigma^2\right)^{1/4}}\ex{-\frac{1}{4\sigma^2}(J_z - m)^2}
\end{equation}

\noindent which is Eq.~(\ref{eq:kraus_map}) in the main text.

\begin{figure}[t!]
    \centering
    \includegraphics[width=1\linewidth]{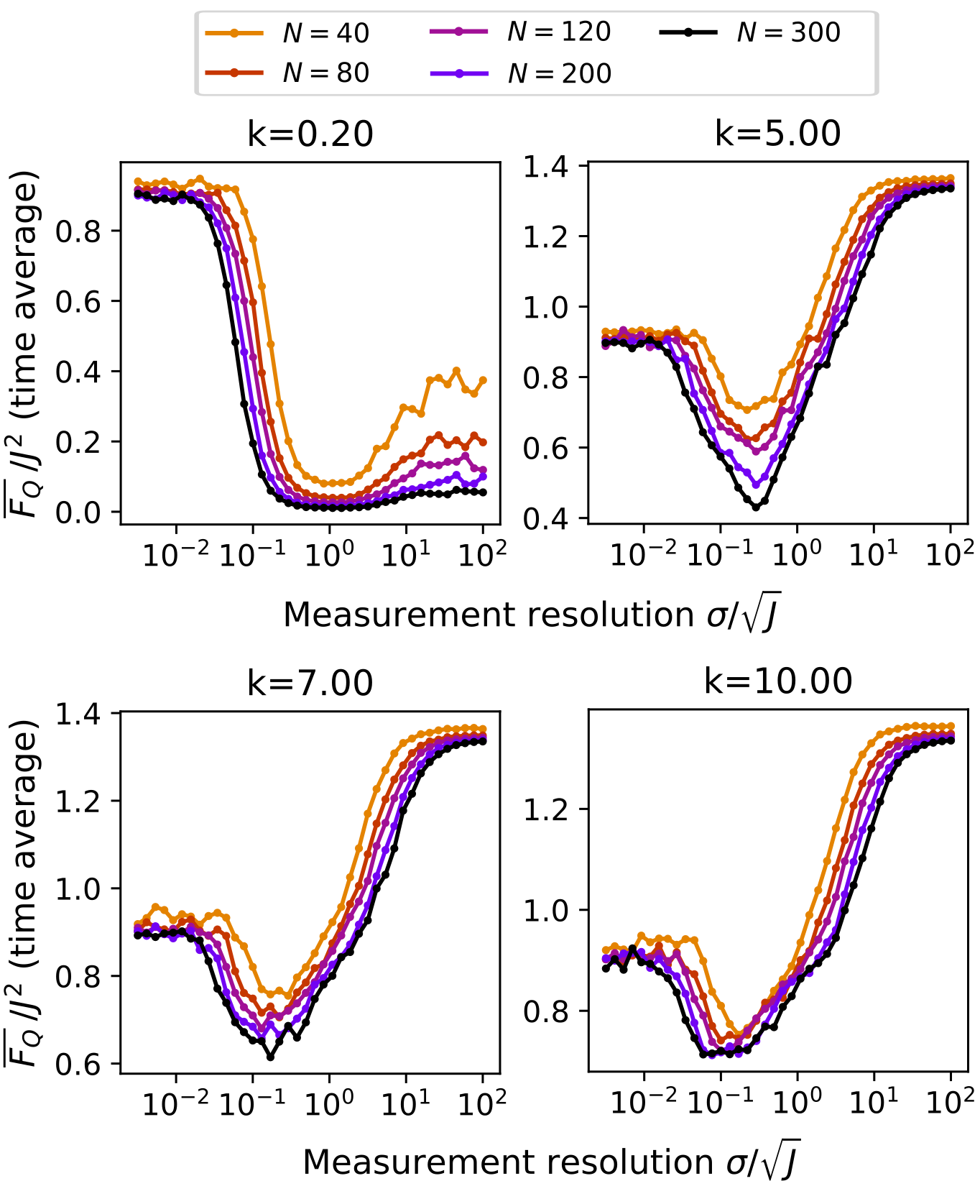}
    \caption{Numerical results obtained using the same procedure as the ones in Fig. \ref{fig:figure2} in the main text, but with different values of the QKT chaoticity parameter $k$. }
    \label{fig:app_fig1}
\end{figure}

\section{Mean QFI for specific states}
\label{app:mean_qfi}

The mean QFI of Eq.~(\ref{eq:avg_qfi}) is proportional to the trace of the covariance matrix $\mathbb{V}$ of the collective magnetization, whose elements are defined as
\begin{equation}
    V_{ij} = \frac{1}{2}\left( \langle \{J_i,J_j\}\rangle - 2\langle J_i \rangle \langle J_j\rangle\right)
\end{equation}

By construction, the trace of $\mathbb{V}$ is independent of the choice of axis $x,y,z$, and obeys
\begin{align}
\trace{\mathbb{V}}= \sum\limits_{\mu=x,y,z} \Delta J_\mu^2 &= \sum_\mu \left( \langle J_\mu^2\rangle - \langle J_\mu\rangle^2\right) \\
    & = J(J+1) - || \langle \mbf{J}\rangle||^2,
\end{align}

\noindent such that the mean QFI is $\overline{F_Q}=\frac{4}{3}\trace{\mathbb{V}}$. We can thus describe this quantity in terms of the properties of the mean magnetization vector. 

For some specific families of states, we can obtain useful analytical expressions for the mean QFI. Spin coherent states (SCSs) are product states that saturate the standard shot-noise limit,
\begin{equation}
    ||\langle \mbf{J}\rangle||^2_{\mathrm{SCS}} = J^2,
\end{equation}
\noindent which follows from the fact that a SCS is the maximum projection eigenstate of the angular momentum operator $J_{\mbf{n}}$ on some direction $\mbf{n}$.\\

States that deviate slightly from exact SCSs can be analyzed via a Gaussian approximation, done through the Holstein-Primakoff transformation with respect to the direction $\mbf{n}$, in which we take
\begin{align}
    J_{\mbf{n}} &= J-n \\
    J_1 &= \sqrt{J}{Q} \\
    J_2 &= \sqrt{J}{P},
\end{align}
with $n$ the number operator of the bosonic mode. Appropriate choice of $\mbf{n}$ here gives $\langle Q\rangle = \langle P\rangle = 0$ by construction, and so the norm of the mean magnetization vector reduces to
\begin{equation}
    || \langle \mbf{J}\rangle||^2 = \langle J_{\mbf{n}}\rangle^2 \simeq J^2 - 2J\langle n\rangle.
\end{equation}

Since $\langle n\rangle = \frac{1}{2}\left(\langle Q^2\rangle + \langle P^2\rangle -1\right)$, we can the expression above in terms of a squeezing parameter $r$ where $\langle Q^2\rangle = e^{-r}/2$ and $\langle P^2\rangle = e^{r}/2$,

\begin{equation}
    || \langle \mbf{J}\rangle||^2_{\mathrm{Gaussian}} = J^2 - J\left(\cosh(r)-1\right)
\end{equation}

Notice that $r$ is typically a function of $N$ (or $J$). For an optimal spin-squeezed state generated through one-axis twisting, we have that $\langle Q^2\rangle \sim N^{-2/3}$~\cite{KitagawaUeda1993,Pezze2018}. 

We can also easily compute the average value of the mean QFI over Haar-random states. From symmetry considerations, this corresponds to three times the average of $\langle J_\mu\rangle^2$ in any direction. We can use a known result
\begin{equation}
    \overline{\langle A\rangle \langle B\rangle} = \frac{1}{d(d+1)}\left( \trace{A}\trace{B} + \trace{AB}\right),
\end{equation}

\noindent and the fact that $\trace{J_\mu^2}=\frac{1}{3}J(J+1)(2J+1)$ to show that
\begin{equation}
    \overline{\langle J_\mu\rangle^2}=\frac{1}{3}\frac{J}{2}\Rightarrow || \langle \mbf{J}\rangle||^2_{\mathrm{Haar}} = \frac{J}{2}.
\end{equation}

Another important case are Dicke states, for which we trivially have that $||\langle \mbf{J}\rangle||^2 = m_z^2$. If $m_z$ is uniform random variable between $-J$ and $J$, we have that

\begin{equation}
    ||\langle \mbf{J}\rangle||^2_{\mathrm{avgDicke}} = \frac{1}{2J+1}\sum\limits_{m_z=-J}^J m_z^2 = \frac{1}{3} J(J+1),
\end{equation}
\noindent which is always larger than the case of Haar random states. The results derived in this Appendix are shown in Table \ref{tabl:avgqfis} of the main text.

\section{Additional numerical results} \label{app:additional_numerics}

\begin{figure}[t!]
    \centering
    \includegraphics[width=1\linewidth]{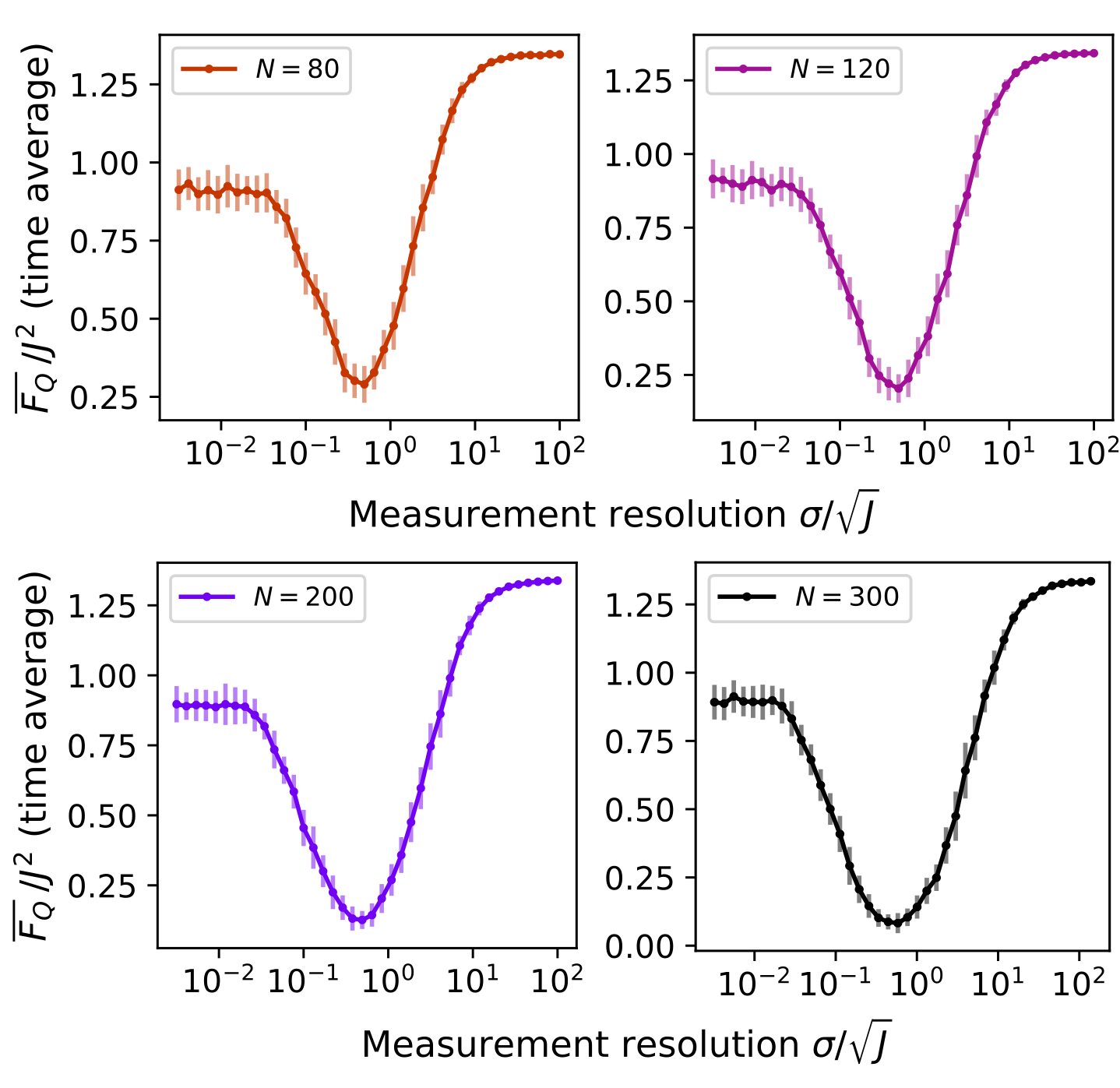}
    \caption{Panels show the same numerical data as Fig. \ref{fig:figure2} (b) (i.e. $k=3$) but now with error bars. These correspond to the standard error of the mean arising from the average over random quantum trajectories. }
    \label{fig:app_fig2}
\end{figure}

Here we expand the presentation of numerical results put forward in Sec.~\ref{sec:results_transition}. First, Fig.~\ref{fig:app_fig1} shows the time-averaged QFI as a function of measurement resolution $\sigma/\sqrt{J}
$ for different system sizes in a similar way as Fig.~\ref{fig:figure2} in the main text. The additional results correspond to other values of the QKT parameter $k$. These range from $k=0.2$, corresponding to a completely regular motion which is almost only rigid rotations, up to $k=10$ which corresponds to the highly chaotic case. Conversely, in Fig.~\ref{fig:app_fig2} we present results only for $k=3$, and each panel corresponds to different system sizes. While the data is exactly the one showed in Fig.~\ref{fig:figure2} in the main text, in this case we show the standard error of the mean for each curve. In all cases we see that error bars are suppressed in the weak measurement regime, where the resulting mean QFI approaches the Haar-random value. Note that this happens for $k=3$ where the system is already chaotic - smaller values of $k$ yield regular motion and present larger error bars typically (not shown here). The opposite limit of strong measurements yield the largest error bars in the typical case, which can be attributed to a larger variance stemming from the uniform Dicke state distribution.\\

\change{Finally, we present some details about the numerical methods used to obtain the results in this paper. All scripts are developed using built-in functions from Numpy 1.21 and Scipy 1.9, and we use Python 3.8. Numerical representation of collective spin states and operators corresponding to $N$ particles can be performed efficiently by restricting to the symmetric subspace, which has dimension $d_S=N+1$ as discussed in Sec. \ref{sec:collective_systems}. All operators in the map of interest, c.f. Eq. (\ref{eq:dyn_model_map}) are represented as matrices in the Dicke basis (i.e., the basis of eigenstates of $J_z$, and we simulate the dynamics of each quantum trajectory exactly by sampling the measurement outcome $m$ for each time step from Eq. (\ref{eq:dyn_model_distro}). As stated in the main text, for each trajectory we store the mean magnetization vector $||\langle \mathbf{J}\rangle||^2$ as a function of discrete time, which in all cases runs up to $M_s=40$. Further details are given in the main text.}

\section{Details of the analytical model of Sec. \ref{ssec:fully_chaotic}} \label{app:analytical_model}

Here we analyze the expression in Eq.~(\ref{eq:haar_resultJ})

\begin{align}
    \mathbb{E}\left[\langle J_{\alpha}\rangle_m^2\right] \simeq 
    \frac{ \trace{K_{m}^2 J_{\alpha}}^2 + \trace{(K_{m}^2 J_{\alpha})^2} }{\trace{K_{m}^2}^2 + \trace{K_{m}^4}},
    \label{eq:haar_resultJ2}
\end{align}

\noindent where $\langle \cdot \rangle_m = \Bra{\Psi_{m}}\cdot\Ket{\Psi_{m}}$. We use this expression to estimate the mean QFI the highly-chaotic case.  First, recall that
\begin{equation}
    K_{m_0} = \frac{1}{\left(2\pi \sigma^2\right)^{1/4}} \sum\limits_{m=-J}^{J} \ex{-\frac{(m - m_0)^2}{4\sigma^2}} \KetBra{m}{m},
\end{equation}

\noindent and let us define
\begin{equation}
    g^{(m_0)}(m)\equiv g(m) = \frac{1}{\sqrt{2\pi \sigma^2}} \ex{-\frac{(m - m_0)^2}{2\sigma^2}}
\end{equation}

\noindent such that $K^2$, which appears a lot in these calculations, is simply
\begin{equation}
    K^2 = \sum\limits_{m=-J}^J g(m) \KetBra{m}{m}.
\end{equation}
\noindent and we have dropped the $m_0$ label to lighten the notation. In the limit of large $J$, we treat $m$ as a continuous variable and get
\begin{widetext}
\begin{align}
    \trace{K^2} &\simeq \int dm\: g(m) = \frac{1}{2}\left( \erf{\frac{J-m_0}{\sqrt{2}\sigma}} +\erf{\frac{J+m_0}{\sqrt{2}\sigma}} \right) \\
    \trace{K^4} & \simeq \int dm\: g(m)^2 = \frac{1}{4 \sqrt{\pi\sigma^2}}\left( \erf{\frac{J-m_0}{\sigma}} +\erf{\frac{J+m_0}{\sigma}} \right).
\end{align}
\end{widetext}

The terms in the numerator depend on the choice of axis. We start with  $A=J_z$,

\begin{widetext}
\begin{align}
    \trace{K^2 J_z} & \simeq \int dm\: m\:g(m) = \frac{\sigma}{\sqrt{2\pi}} \left(e^{- \frac{(J+m_0)^2}{2\sigma^2}} -  e^{- \frac{(J-m_0)^2}{2\sigma^2}}\right) + m_0 \trace{K^2} \\
    \trace{K^2 J_z K^2 J_z} & \simeq \int dm\: m^2\:g(m) = \frac{1}{4\pi} \left( (m_0-J) e^{- \frac{(J+m_0)^2}{\sigma^2}} - (m_0+J) e^{- \frac{(J-m_0)^2}{\sigma^2}} \right) + \nonumber \\
    & + \left(m_0^2+\frac{1}{2}\sigma^2\right)\trace{K^4}.
\end{align}
\end{widetext}

Then, we expect to get the same results for $J_x$ and $J_y$ due to symmetry, since the Kraus operator $K_m$ is invariant under continuous rotation around the $z$-axis. Upfront we have
\begin{subequations}
\begin{align}
    \trace{K^2 J_x} &= \sum\limits_m g(m) \Bra{m}J_x\Ket{m} = 0 \\
    \trace{K^2 J_x K^2 J_x} &= \sum\limits_{l,m} g(l)g(m) \Bra{m} J_x \Ket{l}^2.
\end{align}
\end{subequations}

Since $\Bra{m} J_x \Ket{l}=\frac{1}{2}\left(\delta_{l,m-1} C_m^{-} + \delta_{l,m+1}C_m^+\right)$, where ${C_m^{\pm}}^2 = (J\mp m)(J\pm m+1)$, the expression yields
\begin{widetext}
\begin{align}
    \trace{K^2 J_x K^2 J_x} &= \frac{1}{4}\sum\limits_m\left( g(m-1)g(m) {C_m^-}^2 + g(m+1)g(m){C_m^+}^2\right) \nonumber \\
    &= \frac{1}{4}\int dm\: g(m-1)g(m) \left({C_m^-}^2 + {C_{m-1}^+}^2\right) \nonumber \\
    &\simeq  \frac{1}{2}\int dm\: g(m-1)g(m)\left(J^2-m^2\right).
\end{align}
\end{widetext}
Notice that we have approximated $J+1\simeq J$ in the last step. This integral can actually be solved exactly, the results ultimately takes the form
\begin{widetext}
\begin{align}
    \trace{K^2 J_x K^2 J_x} &= \frac{1}{8\pi}\left\{ (J+m_0)e^{\frac{1}{\sigma^2} \left(-(J^2+m_0^2)-m_0(1-2J)+J\right)} + \right. \nonumber \\
    & \left. (J-m_0)e^{\frac{1}{\sigma^2} \left(-(J^2+m_0^2)-m_0(1+2J)-J\right)} \right\} -\frac{1}{32\sqrt{\pi}\sigma} e^{-\frac{1}{4\sigma^2}}\left\{ -4J^2 +2\sigma^2 \right. \nonumber \\
    & \left. +(1+2m_0)^2\right\}\left( \erf{\frac{J-m_0}{\sigma}} +\erf{\frac{J+m_0}{\sigma}} \right).
\end{align}
\end{widetext}

\bibliographystyle{quantum}
\bibliography{miept_collective}
\end{document}